\documentclass[twocolumn,superscriptaddress,floatfix,preprintnumbers,nofootinbib,prd,10pt]{revtex4-2}
\usepackage{docs}%
\usepackage{bm}%
\usepackage[colorlinks=true,linkcolor=blue, citecolor=red]{hyperref}
\pdfoutput=1
\usepackage{graphicx}
\usepackage{xcolor,colortbl}
\usepackage{subfig}
\usepackage{dcolumn}
\usepackage{mathtools}
\usepackage{bm}
\usepackage{amssymb}
\usepackage{amsmath}
\usepackage{epsfig}
\usepackage[toc,page]{appendix}
\usepackage{slashed}
\usepackage{hhline}
\usepackage{color}
\usepackage{amsmath} 
\usepackage{upgreek} 
\usepackage{bm} 
\usepackage{multirow}
\usepackage{url}
\usepackage{empheq}
\usepackage{times}

\newcommand{\bmt}{\begin{pmatrix}}
\newcommand{\emt}{\end{pmatrix}}
\newcommand{\ba}{\begin{array}{c}}
\newcommand{\ea}{\end{array}}
\newcommand{\be}{\begin{equation}}
\newcommand{\ee}{\end{equation}}
\newcommand{\bea}{\begin{eqnarray}}
\newcommand{\eea}{\end{eqnarray}}
\newcommand{\nn}{\nonumber}
\newcommand{\bi}{\begin{itemize}}
\newcommand{\ei}{\end{itemize}}

\newcommand{\baz}{\begin{array}{cc}}

\begin{document}

\title{Effect of scalar condensation on fermionic pole-skipping}

\author{Banashree Baishya}
\email[E-mail: ]{b.banashree@iitg.ac.in}
\affiliation{Department of Physics, Indian Institute of Technology Guwahati, Assam-781039, India}

\author{Sayan Chakrabarti}
\email[E-mail: ]{sayan.chakrabarti@iitg.ac.in}
\affiliation{Department of Physics, Indian Institute of Technology Guwahati, Assam-781039, India} 

\author{Debaprasad Maity} 
\email[E-mail: ]{debu@iitg.ac.in}
\affiliation{Department of Physics, Indian Institute of Technology Guwahati, Assam-781039, India}

\begin{abstract} 
In this paper, we investigated the holographic fermionic pole-skipping phenomena for a class of interacting theory in a charged AdS black hole background. We have studied two types of fermion-scalar interactions in the bulk: Dipole and Yukawa type interaction. Depending upon the interaction we introduced both real and charged scalar fields. We have particularly analyzed the effect of scalar condensation on the fermionic pole-skipping points and discussed their behaviour near critical temperatures.
\end{abstract}

\maketitle


\section{Introduction}
Strongly coupled interacting field theory is still far from our current understanding. The gauge/gravity correspondence \cite{Maldacena:1997re, Aharony:1999ti, Polchinski:2010hw} is believed to be a well-known approach in understanding such a strongly coupled system. In this framework, strong-weak duality is at the heart of this correspondence where a strongly coupled field theory living on the boundary of AdS space is expected to have unique correspondence with a weakly coupled gravitational theories in AdS space \cite{Gubser:1998bc, Witten:1998qj, Zaffaroni:2000vh}. Such correspondence has been further generalized to a finite temperature quantum system which is dual to an asymptotically AdS black hole. From the field theory perspective, when such a finite temperature equilibrium system has been perturbed, the dynamics of such perturbation are generically encoded in the retarded two-point correlation function \cite{Son:2002sd, Iqbal:2009fd} of the associated dual quantum operators on the boundary. Every quantum operator at the boundary field theory has its dual field in the AdS bulk. When the field theory under consideration is strongly coupled, the retarded correlation function of the boundary operator can be calculated by solving the classical equation of motion of the dual bulk field in the AdS black hole spacetime. The boundary retarded Green's function is uniquely defined by setting ingoing boundary conditions of the field near the black hole horizon in the bulk. Apart from setting natural boundary conditions, the near horizon dynamics of bulk field provide interesting insights into the infrared behaviour of the retarded Green's function in terms of the dispersion relation. Such a relation appeared to have a universal holographic form in the zero temperature limit and is governed by the AdS$_2$ Green's function. However, high-energy behaviour is non-trivially dependent on the radial evolution of the field in full black hole geometry. Recently a striking new observation has been made \cite{Grozdanov:2017ajz, Blake:2017ris} of this Green's function when one moves away from the centre of the complex Fourier $(k,\omega)$ plane. It is observed that for certain discrete sets of complex values of $(k,\omega)$, the retarded Green's function loses its uniqueness, and when it applies to that of the energy-momentum operators, non-uniqueness seems to characterize the chaos of the system. The discrete points in the complex $(k,\omega)$ plane where such phenomena occur are dubbed as pole-skipping points \cite{Blake:2018leo}. These are the points where lines of poles and zeroes of boundary Green's function cross each other. Those special points have further been shown to have an elegant holographic interpretation in terms of the non-unique boundary condition of the bulk metric perturbation near the black hole horizon \cite{Natsuume:2019xcy, Blake:2019otz}. Similar to the pole-skipping phenomenon, this intersection of lines of poles and zeroes occurs in some overdoped metals. Experimental studies on doped cuprates revealed this anomalous behaviour. Generally, the Fermi surface is defined by the poles of Green's function $G(k,\omega)$ at frequency $\omega=0\,$ (near $T_c$). In the Mott insulator, the Fermi surface disappears while the zero surface appears. So, the point of transition from metals to insulators can be thought of as a pole-skipping point, where both poles and zeroes surfaces overlap. However, it's still a point of discussion how these metals transit into the Mott insulator\cite{Sakai_2009}. Studying pole-skipping in holographic systems may give an insight into the transition. The dipole coupling model can be a very good candidate for studying this overlapping phenomenon\cite{Alsup:2014uca}. Subsequently, understanding this pole-skipping phenomenon in the holographic context has been generalized to other fundamental fields namely, scalar field, gauge field \cite{Blake:2019otz, Natsuume:2019sfp}, Fermion field\cite{Ceplak:2019ymw}. Pole-skipping has been extensively studied with finite coupling corrections \cite{Natsuume:2019vcv}, higher curvature corrections \cite{Wu:2019esr} and Scalar Gauss- Bonnet corrections \cite{BAISHYA2024116521}. Pole-skipping has been explored in various contexts in \cite{Kim:2021hqy, Jeong:2021zhz, Ceplak:2021efc, Abbasi:2020xli, Grozdanov:2020koi, Wang:2022mcq, Natsuume:2023lzy, Kim:2021xdz, Sil:2020jhr}.
Boundary field theory interpretations of such phenomena are still under active research. Most of the studies so far have been focused on the free field dynamics in bulk. However, an interacting system is of fundamental importance when we intend to understand the system under study for various non-trivial deformations by introducing non-trivial interaction in the system, on which not many studies have been performed. Keeping this motivation in mind we study in detail the fermionic pole-skipping phenomena under two types of interactions in the bulk, namely, dipole type and Yukawa type coupling. Properties of real-time dynamics due to those interactions have been studied in the literature \cite{Chakrabarti:2019gow, Seo:2018hrc}. In \cite{Chakrabarti:2019gow, Seo:2018hrc}, the authors studied the fermionic spectral function with dipole couplings. But, in our work, we have studied the pole and zero crossing point of the Green's function. We investigate the influence of those coupled specifically sourced by additional scalar fields which can play the role of both source and condensation in the boundary. 

 We have organized the paper as follows: In the first section, we have given a brief introduction to pole-skipping(P-S). In section \eqref{sec-2}, we have discussed the behaviour of the scalar field (both real and charged) which is coupled to dipole and Yukawa couplings. In section \eqref{sec-3}, we have discussed the P-S analysis and shown some non-trivial effects of these couplings on P-S points. Finally, we have summarised the paper with some future directions in section \eqref{sec4}.   
\section{Set-up}\label{sec-2}
Pole-skipping is a finite temperature phenomenon. Therefore, we consider an asymptotically AdS background with a charged black hole in the bulk. As already stated in the introduction, extensive investigations have been done on this phenomenon in various black hole backgrounds considering different fundamental fields. In our present study, we particularly focus on investigating the influence of different interactions in the phenomena of pole-skipping. For our study we consider charged fermions propagate in the Reissner-Nordstr$\Ddot{\text{o}}$m-AdS$_{4}$ black hole background with different coupling terms such as dipole coupling and Yukawa coupling. We further assume that those couplings are facilitated by a bulk scalar, for which we consider both neutral and charged scalar fields. We investigate the non-trivial effect of different scalar field configurations on fermionic pole-skipping phenomena. We will solve the scalar field in RN-AdS$_4$ background and this geometry has a metric as (setting AdS radius $L=1$),
\begin{equation}
    \label{met} dS^{2}={r^{2}}\left[-f(r)dt^{2}+d{x}^{2}+d y^2\right]+\frac{1}{r^2 f(r)}dr^2
\end{equation}
The emblackening factor $f(r)$ and gauge field at the horizon $r_0=1$ is,
\begin{align}
f(r)=1+\frac{3\eta}{r^4}-\frac{1+3\eta}{r^3},\quad
A_{t}=\mu \left(1-\frac{1}{r}\right)dt
\end{align}
We have treated $\frac{Q^2}{3}$ as $\eta$, where $Q$ is the charge of the black hole and $\mu$ is the chemical potential. The temperature of the Black hole is $T=\frac{3}{4\pi}(1-\eta)$ and we will work in the range $0<\eta<1$. 
\subsection{\textbf{Real scalar field}}
We consider the action that coupled to gravity in AdS$_4$ with real massive scalar field $\Phi$ is,
\begin{align}
   \nn\mathcal{S}&=\frac{1}{2{\mathcal{K}}^2}\int d^{4}x\sqrt{-g}\left[\mathcal{R}+\frac{6}{L^2}-\frac{1}{4}F^{2}\right.\\&\hspace{2cm}\left.+\frac{1}{\lambda}\left( -\frac{1}{2}g^{\mu\nu}\nabla_{\mu}\Phi\nabla_{\nu}\Phi-V(\Phi)\right) \right] .
\end{align}
Where $L$ is the AdS radius and $\lambda$ is the coupling constant. Later on, we will set $L=1$ for our convenience and $\lambda$ to be very large. We choose the potential as,
\begin{equation}
V(\Phi)=\frac{1}{4}\left(\Phi^2+m_{\Phi}^{2} \right)^{2}-\frac{m_{\Phi}^4}{4}.
\end{equation}
We can easily get the scalar field equation as
\begin{equation}
\frac{1}{\sqrt{-g}}\nabla_{\mu}\left(\sqrt{-g}g^{\mu\nu}\nabla_{\nu}\Phi \right)-\left(\Phi^2+m_{\Phi}^2\right)\Phi=0\,.
\end{equation}
Now, we will shift our co-ordinates to ingoing Eddington-Finkelstein co-ordinates by making the following  transformation,
\begin{equation}
v=t+r_*,\hspace{1cm}\frac{dr_*}{dr}=\frac{1}{r^2f(r)},\hspace{1cm}r_{0}^2f'(r_0)=4\pi T ,
\end{equation}
where $T$ is the Hawking temperature.
In the new coordinate, the metric and gauge connection are expressed as,
\begin{align}\label{EF}
dS^{2}&=-r^{2}f(r)dv^{2}+2dvdr+r^{2}(dx^{2}+dy^{2}),\\
A&=\mu\left[1-\left( \frac{r_0}{r}\right)\right]dv .
\end{align}

Considering $\Phi(r,v,x,y)=\phi(r)$, the Klein-Gordon (K-G) equation becomes,
\begin{equation}\label{z} 
\phi''(r)+\left(\frac{f'(r)}{f(r)}+\frac{4}{r} \right)\phi'(r)-\frac{(\Phi^{2}(r)+m_{\phi}^2)}{r^{2}f(r)}\phi(r)=0\,.
\end{equation}
The above equation is the radial part of the whole K-G equation, where $\phi(r)$ is the radial part. As we want to solve the radial equation, we have not considered the time and spatial part. Expanding the equation $\eqref{z}$ upto first order at the horizon, yields the following equation,
 \begin{equation}
 \phi'(r_0)=\frac{\left(\phi(r_0)^{2}+m_{\Phi}^{2}\right)\phi(r_0)}{r_{0}^{2}f'(r_0)}\,.
 \end{equation}
The asymptotic (at $r\rightarrow\infty$) behaviour of equation $\eqref{z}$ gives,
\begin{equation}\label{zz} 
\lim_{r\rightarrow\infty}\phi(r)=\mathcal{O}_{1}r^{\Delta-3}+\mathcal{O}_{2}r^{-\Delta}\,,
\end{equation}
 where, at infinity (where is our boundary), normalizable co-efficient  $\mathcal{O}_{1}$ is generically identified as a source and non-normalizable co-efficient $\mathcal{O}_{2}$ is the condensation and scaling dimension of operator, $\Delta=\frac{3}{2}+\sqrt{\frac{9}{4}+m_{\Phi}^{2}}$ is related to mass of scalar, which must satisfy Breitenlohner and Freedman (BF) bound $-\frac{9}{4}<m_{\Phi}^2$ \cite{Breitenlohner:1982bm}. If we consider the mass within specific range  $-\frac{9}{4}<m_{\Phi}^2<-\frac{5}{4}$, there exist two different AdS-invariant quantization schemes \cite{Breitenlohner:1982jf}. One Lagrangian can give rise to two theories in AdS space depending upon the scheme. Both $\mathcal{O}_1$ and $\mathcal{O}_2$ are normalizable solutions\cite{Klebanov_1999} with this given $m_{\Phi}^2$. So, we can treat any of them as a condensate of the field theory operator depending upon the scheme.
From equation $\eqref{zz}$, we can write,
\begin{equation}\label{z1} 
\lim_{r\rightarrow\infty}r\phi'(r)=(\Delta-3)\mathcal{O}_{1}\,r^{\Delta-3}-\Delta\mathcal{O}_{2}\,r^{-\Delta}\,.
\end{equation}
Solving the K-G equation near the boundary, we can easily write $\mathcal{O}_1$ and $\mathcal{O}_2$ as,

\begin{subequations}\label{o1o2}
\begin{align}
\mathcal{O}_{1}&=\lim_{r\rightarrow\infty}\frac{r^{3-\Delta}\left(\Delta\phi(r)+r\phi'(r) \right) }{2\Delta-3},\\
\mathcal{O}_{2}&=\lim_{r\rightarrow\infty}\frac{r^{\Delta}\left((\Delta-3)\phi(r)-r\phi'(r) \right) }{2\Delta-3}.
\end{align}
\end{subequations}

\begin{figure*}[t]
\centering
\includegraphics[height=0.27\textwidth]{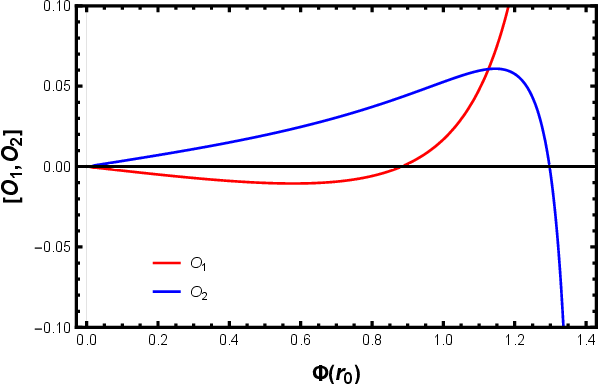}
\includegraphics[height=0.27\textwidth]{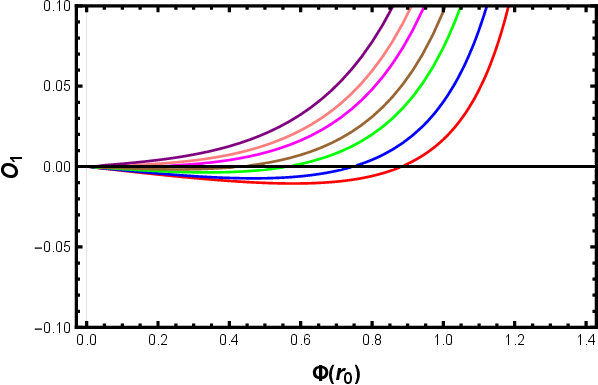}
\caption{\textit{Left:} Behaviour of $\mathcal{O}_{1}$ and $\mathcal{O}_{2}$ with increasing horizon value of real scalar field at T=0.0002. \textit{Right:} Behaviour of $\mathcal{O}_{1}$ with horizon value of real scalar field by varying temperature. The temperature values from the purple curve to the red curve are (0.0016, 0.0012, 0.001, 0.0007, 0.0005, 0.0003, 0.0002) respectively. We have fixed $m_{\Phi}^{2}=-2.1$ in both the plots. The critical temperature $T_c$ has been calculated from the curve beyond which it does not touch the horizontal axis.}
\label{scal_plot}
\end{figure*}

\begin{figure*}[t]
\centering
\includegraphics[height=0.29\textwidth]{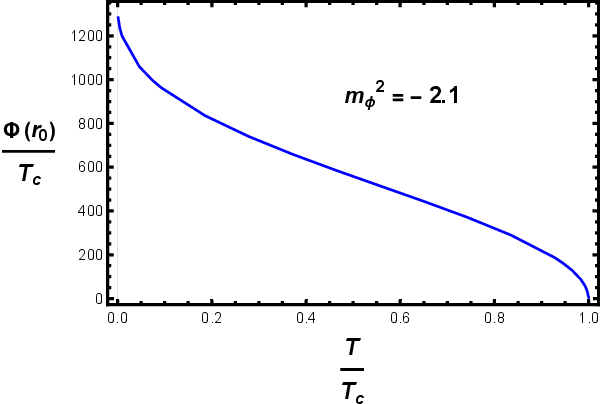}
\includegraphics[height=0.29\textwidth]{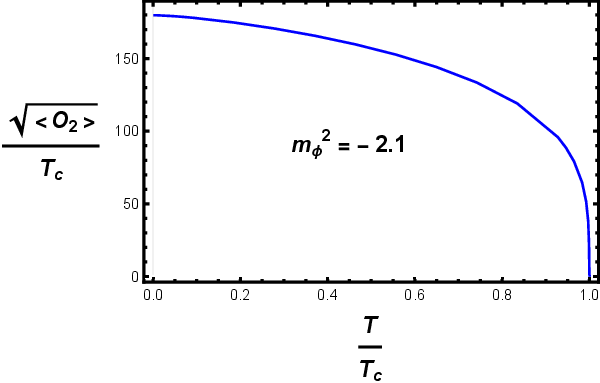}
\caption{\textit{Left:}Behaviour of $\phi(r_0)$ with increasing temperature. \textit{Right:} Behaviour of $\sqrt{\mathcal{O}_{2}}$  by varying temperature. Both the plots are for real scalar fields. We see that at a certain value of temperature, the horizon value of the field vanishes and the condensate also vanishes at the same temperature. $<.>$ means the vacuum expectation value of the operator.}
\label{scal_plot2}
\end{figure*}

Now, choosing the standard scheme of quantization, we can treat $\mathcal{O}_{1}$ as the source and $\mathcal{O}_{2}$ as condensate. We derive the equation of motion for the scalar field and impose regularity at the horizon to establish the boundary conditions. Using the shooting method, we then solve for the scalar field from the near-horizon region to the boundary by appropriately choosing the mass of the scalar field ($m_\Phi$), the temperature, and the scalar field value at the horizon ($\phi(r_0)$). Once the numerical solution is obtained we compute $\mathcal{O}_1$ and $\mathcal{O}_2$ from eq. \eqref{o1o2} as depicted in the left panel of \figurename{\ref{scal_plot}} at T=0.0002. The right panel of \figurename{\ref{scal_plot}} is generated by varying the temperature to study the source behaviour. We can see from the right plot of the \figurename{ \ref{scal_plot2}} that as we increase the temperature, after a critical temperature the horizon value turns to zero. With the given parameters $m_{\phi}^2=-2.1$, the critical temperature is obtained as $T_c=0.0011$. Near this critical temperature $T_c$, we have fitted $\phi(r_0)$ with  $\gamma(T_c-T)^\delta$, which gives the exponent value $\delta\approx 0.56\pm 0.009$ and $\gamma\approx 40.91$. The exponent is close to the mean field value $1/2$. Near $T=T_c$, to the leading order we can assume $\phi = \epsilon^\delta (\psi(r) + {\cal O}(\epsilon)) $ with $\epsilon = T_c-T$, and $\delta >0$. For the phase transition to occur, it is the $\phi^2$ term that plays an important role in equation \eqref{z}. Since near $T=T_c$, the coefficients of $\phi''$ and $\phi'$ are polynomials in $\epsilon$. The coefficient $\phi^2$ should also be polynomial in $\epsilon$ for consistency and hence $2\delta$ should be integer with its lowest value being $\delta = 1/2$.    

\subsection{\textbf{Charged scalar field}}\label{charged_sec}
Now, we consider a minimally coupled charged scalar field $\Tilde{\Phi}$ in Reissner-Nordstr$\Ddot{\text{o}}$m-AdS$_{4}$ (RN-AdS$_4$) black hole space-time ,
\begin{align}
   \nn\mathcal{S}&=\frac{1}{2{\mathcal{K}}^2}\int d^{4}x\sqrt{-g}\left[\mathcal{R}+\frac{6}{L^2}-\frac{1}{4}F^{2}\right.\\&\hspace{2cm}\left.+\frac{1}{\lambda}\left( |\partial\Tilde{\Phi}-iq_{s}A\Tilde{\Phi}|^2-V(\Tilde{\Phi})\right) \right]\,.
\end{align}
Again we will consider $\lambda$ to be very large so that the charged scalar does not back-react on the background geometry. Here, $q_s$ is the charge of the scalar field and $A$ is the gauge field. We will work with the same potential as in the case of the real scalar field, but $\phi$ is replaced by $|\Tilde{\Phi}|$.
Assuming the ansatz $\Tilde{\Phi}=\Tilde{\phi}(r)$, we get the K-G equation as
\begin{align}\label{charged_eom}
  & \Tilde{\phi}''(r)+\left(\frac{f'(r)}{f(r)}\right)\Tilde{\phi}'(r)-\frac{1}{r^{2}f(r)}\left((m^{2}+\Tilde{\phi}^2(r))\right.\nn\\&\hspace{3cm}\left.-q_{s}^{2}A^{2}(r)f(r)\right)\Tilde{\phi}(r)=0\,.
\end{align}
Solving the equation of motion (EOM) near the AdS boundary, we can calculate the source and condensate value which gives the same form as \eqref{o1o2}. The asymptotic behaviour of this EOM gives the same form as equation \eqref{zz}. So, with the same approach as in the case of the real scalar field, we can calculate $\mathcal{O}_1$ and $\mathcal{O}_2$ numerically for the charged case also.

\begin{figure*}[t]
\begin{minipage}[b]{0.41\linewidth}
\centering
\includegraphics[width=\textwidth]{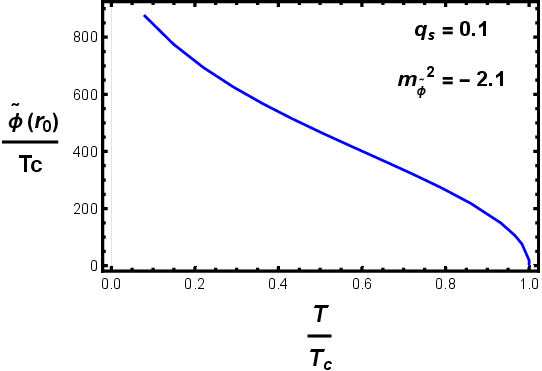}
\end{minipage}
\hspace{0.4cm}
\begin{minipage}[b]{0.45\linewidth}
\centering
\includegraphics[width=\textwidth]{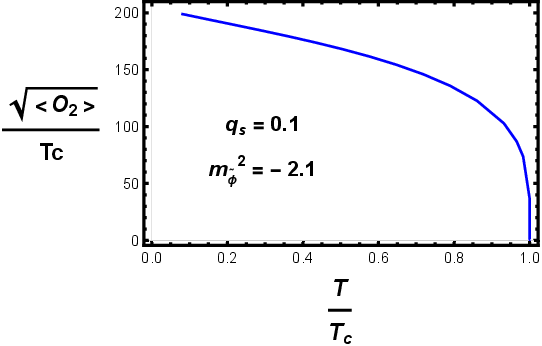}
\end{minipage}
\caption{\textit{Left:} Behaviour of horizon value of the scalar field with temperature. \textit{Right:} Behaviour of $\sqrt{\mathcal{O}_{2}}$ with temperature. Both the plots are for charged scalar fields. At a certain value of temperature $T_c$, we see that the horizon value of the charged scalar field and charged scalar condensate vanishes.}
\label{charged_scal_plot}
\end{figure*}

In \figurename{\ref{charged_scal_plot}}, we choose to work in standard quantization. For the given mass  $m_{\Tilde{\phi}}^2=-2.1$ and charge $q_s= 0.1$ the associated critical temperature comes out to be $T_c=0.0014$. 
We can see that at critical temperature $T_c$, both the horizon value of the scalar field and condensate become zero. Near this critical temperature $T_c$, we have fitted $\tilde{\phi}(r_0)$ with  $\alpha(T_c-T)^\beta$, which gives the exponent value $\beta\approx 0.58\pm 0.01$ and $\alpha\approx 45.38$. Here the exponent is close to the mean field value which is expected.

\section{Pole-skipping analysis}\label{sec-3}
For the pole-skipping phenomenon, we will be working in ingoing Eddington-Finkelstein co-ordinates \eqref{EF}. Since our metric is not diagonal, we choose a frame given by,

\begin{align}
  \nn E^{\underline{v}}&=
\dfrac{1+f(r)}{2}rdv-\dfrac{dr}{r};\hspace{0.5cm}
E^{\underline{r}}=
\dfrac{1-f(r)}{2}rdv+\dfrac{dr}{r} ,\\
&\hspace{2cm}E^{\underline{x}}=
r dx;\hspace{0.5cm}
E^{\underline{y}}=
r dy ,
\end{align}
for which 
\begin{align*}
ds^{2}&=\eta_{\underline{ab}}E^{\underline{a}}E^{\underline{b}} & \eta_{\underline{ab}}&=diag(-1,1,1,1).
\end{align*}
The underlined co-ordinates are assumed to be flat tangent space co-ordinates and the non-underlined are the spacetime co-ordinates. The Gamma matrices in appropriate basis are assumed as,
\begin{eqnarray}
\Gamma^{\underline{r}}&=\begin{pmatrix}
\mathbb{I} & 0
\\0 & -\mathbb{I}
\end{pmatrix},\hspace{1cm}
\Gamma^{\underline{v}}&=\begin{pmatrix}
0&i\sigma^{2}\\i\sigma^{2}&0
\end{pmatrix}\\
\Gamma^{\underline{x}}&=\begin{pmatrix}
0 &\sigma^{1}
\\\sigma^{1}&0
\end{pmatrix},\hspace{1cm}
\Gamma^{\underline{y}}&=\begin{pmatrix}
0&\sigma^{3}\\\sigma^{3}&0
\end{pmatrix} .
\end{eqnarray}
In the above-written Gamma matrices, $\sigma$'s are the Pauli matrices. We consider the interacting charged Fermion Lagrangian as
\begin{equation}\label{lagra} 
\mathcal{L}=\sqrt{-g} i\bar{\Psi}(\slashed{D}-m+\zeta(\phi))\Psi
\end{equation}
Where,\begin{align}
\slashed{D}&=e^{M}_{\underline{c}}\Gamma^{\underline{c}}(\partial_{M}+\frac{1}{4}\omega_{\underline{ab}M}\Gamma^{\underline{ab}}-iqA_{M})\,,
\end{align}
here, $M$ is space coordinate, $a,b,c$ are flat coordinates, $e^{M}_{\underline{c}}$ are vielbein and $\omega_{\underline{ab} M}$ are spin connection terms. $\Gamma^{\underline{ab}}$ is a Gamma matrix that will follow Clifford algebra. Now, in this section, we will calculate the P-S points for the following fermionic coupling prescription, $\zeta(\phi)$ as
\begin{itemize}
    \item When $ \zeta(\phi)=-ip\phi\slashed{F}$: The interaction term $-ip\bar{\psi}\phi\slashed{F}\psi$ is known as dipole coupling where the gauge field  $A_\mu$ is coupled with  a real scalar field $\phi(r)$ with a coupling parameter $p$. Here, $\slashed{F}=\frac{1}{2}\Gamma^{\underline{ab}}e^{M}_{\underline{a}}e^{N}_{\underline{b}}F_{MN}$ and $F_{MN}$ is the electromagnetic field strength.
    \item When $ \zeta(\phi)=g\phi$: The interaction term $g\bar{\psi}\phi\psi$ is the Yukawa coupling term and $g$ is the coupling parameter.
\end{itemize}
The equation of motion for $\Psi$ is,
\begin{equation}\label{eom} 
(\slashed{D}-m+\zeta(\phi))\Psi=0 .
\end{equation}
In the momentum space, we decompose $\Psi=\psi(r)e^{-i\omega v+ik_{x}x}$ setting $k_y=0$ using the rotational symmetry in the x-y plane. So, we will write $k_x=k$ in our calculations. With this equation \eqref{eom} becomes,
 \begin{align}
     & \nn\left[\Gamma^{\underline{v}}\left[-\frac{r}{2}(1-f(r))\partial_{r}-\frac{i\omega}{r}-\frac{iqA_v}{r}-\frac{3}{4}(1-f(r))+\frac{rf'(r)}{4}\right]\right.\\\nn&\left.\quad+\Gamma^{\underline{r}}\left[\frac{r}{2}(1+f(r))\partial_{r}-\frac{i\omega}{r}-\frac{iqA_v}{r}+\frac{3}{4}(1+f(r))+\frac{rf'(r)}{4}\right]\right.\\&\left.\quad\quad\quad\quad\quad\quad\quad\quad\quad\quad\quad+\frac{ik}{r}\Gamma^{\underline{x}}-m+\zeta(\phi)\right]\psi=0\,.
   \end{align}
As the matrix $\Gamma^{\underline{r}}$ has two eigen values $1$ and $-1$, we can decompose $\psi$ into two spinor components as $\psi_{+}$ and $\psi_{-}$. Further, we can introduce another decomposition for $\psi_{\pm}$ as the two matrices $\Gamma^{\underline{r}}$ and $k_{x}\Gamma^{\underline{vx}}$ are independent and commuting. Below we have written the decompositions as
   \begin{align}
       \psi&=\psi_{+}+\psi_{-},\quad\quad\Gamma^{\underline{r}}\psi_{\pm}=\pm\psi_{\pm},\quad\quad\Gamma^{\underline{r}}\Gamma^{\underline{a}}\psi_{\pm}=\pm\Gamma^{\underline{a}}\psi_{\pm}\\
       \psi_{+}&=\psi_{+}^{+}+\psi_{+}^{-},\quad\quad\psi_{-}=\psi_{-}^{+}+\psi_{-}^{-}\quad\quad k_{x}\Gamma^{\underline{vx}}\psi_{\pm}^{\pm}=\pm k\psi_{\pm}^{\pm}
   \end{align}
Now, we can decompose the spinor $\psi$ into $4$ spinor components $(\psi_{+}^{+},\psi_{+}^{-},\psi_{-}^{+},\psi_{-}^{-})$ and get $4$ coupled Dirac component equations; where, $a=v,x,y$.  Then, expanding the Dirac component equations near the horizon order by order, we can calculate the P-S points in each order as shown in detail in the appendix.
\subsection{\textbf{ P-S with real scalar coupling}}
 As we have discussed in the previous section, the coupled scalar can be real or charged. In this section, we will discuss the effect of condensate on P-S points taking the scalar to be real in equation \eqref{eom}. Zeroth order P-S points are found to be,
\begin{align}\label{ps_real}  
\omega_{0}=-\pi iT,\hspace{0.2cm}k_{0} &=\begin{cases}
   \pm({imr_{0}+p\mu\phi(r_0)}),\quad\text{Dipole coupling}\\\pm({imr_{0}+igr_{0}\phi(r_0)}),\text{Yukawa coupling} .
\end{cases}
\end{align}
Point to notice that for both the couplings, the pole-skipping points receive the modification only in the linear momentum sector not in the frequency sector. If we don't consider the coupling terms in the action, we have purely imaginary momentum in the pole-skipping points. However, it seems that by adding interaction to the theory, the momentum values achieve a correction term leaving the frequency unchanged! Similar studies have been done in anisotropic plasma (\cite{Sil:2020jhr}), where they have studied quantum chaos by pole-skipping by perturbing the metric. They got a complex momentum and justified that while the imaginary value of the momentum follows the dispersion curve for momentum diffusion, the real part puts a constraint. This means that the imaginary part of the momentum is the diffusive mode and the real part is the propagating mode. However, it is interesting to observe that due to dipole interaction, the fermionic perturbation acquires a real momentum $\propto \mu\phi(r_0)$ at the pole-skipping point which is temperature-dependent. On the other hand, the Yukawa interaction induces a temperature-dependent spatially growing/decaying mode which is $\propto g \phi(r_0)$. We will observe the same effect in the charged scalar-fermion interaction as well.

Here, $(\omega_0,k_0)$ is the zeroth order frequency and momentum. The 1st-order P-S points are $\omega_1=-3\pi iT$ and 3 associated values of momentum as shown in the appendix. For every order $n$, we get $2n+1$ values of momentum.  
Particularly, the noteworthy finding of the present analysis is the scaling behaviour of the pole-skipping momentum near the phase transition point as
 \begin{align}\label{ps_real1}  
k_0 =\begin{cases}
    \pm({imr_{0}+ 40.9 p\mu (T_c-T)^{\frac12}}),\quad\text{~~Dipole coupling}\\\pm({imr_{0}+40.9 ig r_{0} (T_c-T)^{\frac12}}),~~\text{Yukawa coupling}.
    \end{cases}
\end{align}
The scaling exponent turns out to be again $1/2$ as expected from the background condensation.
The last equality corresponds to the P-S point near the critical temperature. At higher order too, this equality holds, which we can see from the momentum values written in the appendix. We can see that near $T_c$, the P-S points vanish which is the same scenario as arises in \cite{Alsup:2014uca}. In \cite{Alsup:2014uca}, they have discussed how with addition of the coupling term resists the coincidence of lines of poles and zeroes. But, this case is true for $T\rightarrow 0$ temperature. In our paper, we have shown that for massless fermions, this is indeed true near $T_c$. 
 
\subsubsection{Effect of real scalar Condensate:}\label{subsub3.1.1}
From \eqref{zz}, we can recognize the co-efficient of slow fall off as the source term $\mathcal{O}_1$ and co-efficient of fast fall off as the condensate term (response) $\mathcal{O}_2$ of the system. Critical temperature $T_c$ is that temperature below which $\mathcal{O}_1=0$, sourceless condition. In the right of the \figurename{ \ref{scal_plot}} we have plotted the dependence of $\mathcal{O}_1$ in terms of the horizon value of the scalar field. Upon increasing the temperature, the curves are approaching towards the origin signifying the existence of critical temperature $T_c$ at which condensate vanishes. Once we obtained the critical temperature we plotted the condensate as well as the horizon value of condensed scalar field $\phi(r_0)$ with increasing temperature as shown in  \figurename{ \ref{scal_plot2}}. 

\begin{figure*}
\centering
\includegraphics[height=0.29\textwidth]{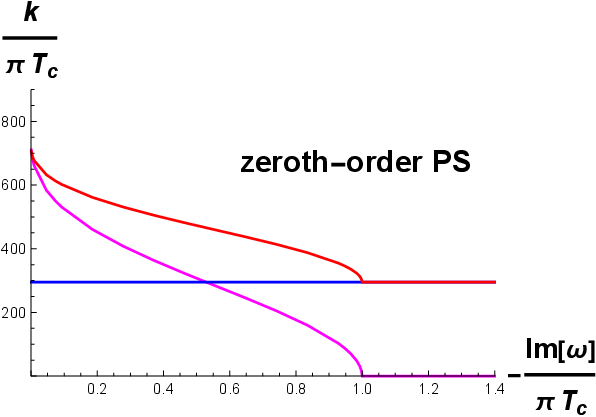}
\includegraphics[height=0.29\textwidth]{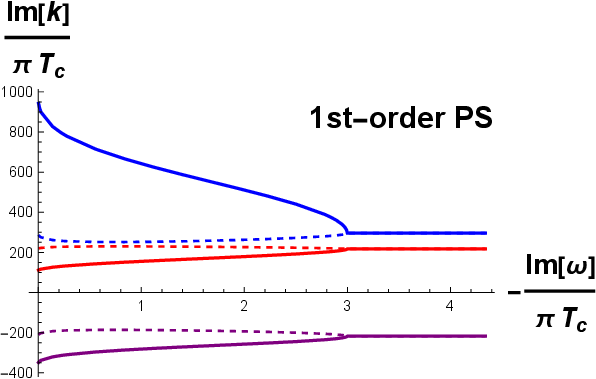}
\caption{\textit{Left:} Movement of zeroth order PS with temperature. The magenta and blue coloured lines are for real and imaginary values of momentum with dipole coupling, while the red colour line is for the imaginary value of momentum with Yukawa coupling. \textit{Right:} Movement of first order pole-skipping point with temperature. Here, we have considered $p=1, g=1, m=1, m_\phi^2=-2.1$ and $q=1$. The thick lines are for momentum values with Yukawa coupling and the dashed lines are for momentum values with dipole coupling. In both the plots, $\mu$ is also varying as $\sqrt{3\eta}$.}
\label{yuk_dip_real_plot}
\end{figure*}

Our goal for this paper is to understand the effects of this scalar condensation on the pole-skipping points. From equation $\eqref{ps_real}$, we can see the horizon value of the scalar field is affecting the pole-skipping point. We have plotted the dynamics of pole-skipping points with varying temperatures in \figurename{ \ref{yuk_dip_real_plot}}. And, we observed that as the temperature increases movement of the pole-skipping point follows the same behaviour as the horizon value of the scalar field. Finally, as the critical temperature is reached, associated momentum saturates with the value which is obtained without interaction. It was expected that beyond the critical temperature, condensate vanishes without any source term. For completeness, we have plotted both zeroth and first-order pole-skipping points. 

\subsection{\textbf{ P-S with charged scalar coupling}}\label{charged_scalar_sec}
\begin{figure*}[t]
\centering
\includegraphics[height=0.29\textwidth]{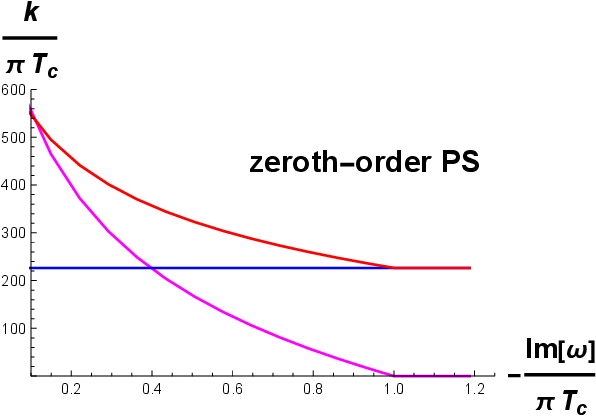}
\includegraphics[height=0.29\textwidth]{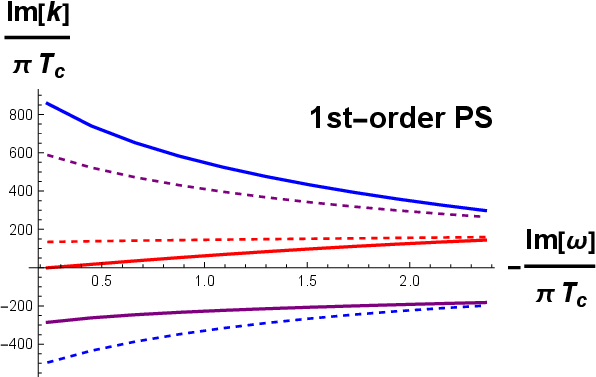}
\caption{\textit{Left:} Movement of zeroth order PS with temperature. The magenta and blue coloured lines are for real and imaginary values of momentum with charged dipole coupling, while the red colour line is for the imaginary value of momentum with charged Yukawa coupling.\textit{Right:} Movement of first order pole-skipping point with temperature. Here, we have considered $p=1, g=1, m=1, m_\phi^2=-2.1,q=1$ and $q_s=0.1$. The thick lines are for momentum values with charged Yukawa coupling and the dashed lines are for momentum values with charged dipole coupling. In both the plots, $\mu$ is also varying as $\sqrt{3\eta}$.}
\label{charged_PS}
\end{figure*}
As we have discussed in the previous section, we can carry the same discussion with charged scalar field and study pole-skipping phenomena. 
\begin{equation}\label{lagrac} 
\mathcal{L}=\sqrt{-g} i\bar{\Psi}(\slashed{D}-m+\zeta(|\tilde{\phi}|^{2}))\Psi\,,
\end{equation}
where, $\tilde{\phi}$ is the charged scalar field. We have already analysed the behaviour of this charged scalar field in detail in section \eqref{charged_sec}. Here also, we can talk about dipole and Yukawa coupling as discussed in the previous section, but with the charged scalar coupling.
\begin{itemize}
    \item when $\zeta(|\tilde{\phi}|^{2})=-ip|\tilde{\phi}|^{2}\slashed{F}$:  This is charged dipole coupling and will see the effect of this interaction term in P-S points.
    \item when $\zeta(|\tilde{\phi}|^{2})=g|\tilde{\phi}|^{2}$: This is charged Yukawa coupling and will see the effect of this interaction term in P-S points.
\end{itemize}
We will follow the same procedure as discussed in the appendix to calculate the P-S points. Zeroth order P-S points are,
\begin{align}\label{ps_charged}  
\omega_{0}=-\pi iT,\hspace{0.1cm}k_{0}=\begin{cases}
    \pm({imr_{0}+p\mu|\tilde{\phi}(r_0)|^2}),\text{ch. Dipole}\\\pm({imr_{0}+igr_{0}|\tilde{\phi}(r_0)|^2}),\text{ch. Yukawa}.
\end{cases}
\end{align}
Here, again $(\omega_0,k_0)$ is the zeroth order frequency and momentum. We have explicitly written down the first-order momentum value in the appendix. Particularly, the noteworthy finding of the charged scalar analysis is the scaling behaviour of the pole-skipping momentum near the phase transition point as
 \begin{align}\label{ps_charged1}  
k_0 =\begin{cases}
    \pm({imr_{0}+ (45.38)^2 \,p\mu (T_c-T)}),\text{ch. Dipole coup.}\\\pm({imr_{0}+ (45.38)^2 \,ig r_{0} (T_c-T)}),\text{ch. Yukawa coup.},
    \end{cases}
\end{align}
The scaling exponent turns out to be unity instead of $1/2$ of the real scalar field. For the first order too, we have verified this behaviour. It is very clear from the \figurename{ \ref{charged_PS}} too. After reaching the critical frequency (where $T=T_c$ ), momentum values saturate at the mass value of fermion.

\subsubsection{Effect of complex scalar Condensate:}
Solving the complex scalar EOM \eqref{charged_eom} near the boundary of AdS, we have evaluated the source and condensate term for the complex scalar field in section \ref{charged_sec}. For this case, we also find the critical temperature $T_c$, beyond which the charge scale becomes trivially zero in source-less condition. Once we obtain the scalar condensate at different temperatures below $T < T_c$, \figurename{ \ref{charged_PS}} depicts the effect of condensation on the zeroth and first order P-S points movement with increasing temperature. After crossing the critical temperature, the P-S points saturate at every order. For massless fermions, P-S points vanish after critical temperature, which seems very interesting. In the \figurename{ \ref{charged_PS}}, we have plotted both zeroth and first-order P-S points for both couplings. For Yukawa coupling, the zeroth order momentum is purely imaginary, so the effect of the coupling terms is additive to the mass of the fermions. But, for dipole coupling, the momentum value is complex. So, we have plotted both real and imaginary values of momentum of dipole coupling in the left of \figurename{ \ref{charged_PS}}. For the first-order P-S points movement, we have only shown the movement of the imaginary values of momentum for both couplings.

\section{Discussion}\label{sec4}
In this paper, we have studied the P-S phenomenon with 2 types of couplings: dipole and Yukawa, where we have probed a scalar field, which can be real or charged. From the knowledge of AdS/CFT, we know that the scalar field near the AdS boundary admits both source and condensate with some specific range of mass values (obeying BF bound). There exists a temperature when this source is zero, still, condensate is finite, which is called the critical temperature. Calculating the P-S points with various couplings as we have mentioned, we have studied the behaviour of these P-S points with increasing temperature up to the critical temperature ($T_c$). We have seen that after crossing $T_c$, the P-S points saturate at the mass value of fermion. For massless fermions, P-S points vanish after crossing $T_c$, which seems very interesting. For real scalar coupling, the momentum values are $\propto\,(T_c-T)^{1/2}$, while for charged scalar coupling, the momentum values are $\propto\,(T_c-T)$. This result holds true in every order and with both couplings, which we have verified in $\textit{mathematica}$ (shown the first order result in appendix). It would be very interesting to see the effect of couplings in other hairy black holes. 

What does it mean when the line of poles and zeros intersect for a real condensed matter system? The poles of the green's function typically define the Fermi surface. An interesting example is the well-known condensed matter system Mott insulator. In those systems, when the system undergoes from metal to insulator transition such phenomenon of pole-skipping happens. At the transition point, fermionic Green's function simultaneously has pole and zeros, where the Fermi surface disappears and zero surface appears\cite{Sakai_2009}. \par In this paper, we have tried to interpret the meaning of pole-skipping points in a novel way by perturbing the fermions. Pole-skipping points are the points of intersection of lines of poles and zeroes of Green's function. In a condensed matter system, a line of poles means the appearance of the Fermi surface, while a line of zeroes means Green's function is zero. The fermionic Green's function is zero means the system is not giving any response to the fermionic perturbations. Or, in a mathematical sense, we can say that the density of states is zero, meaning there are no available energy states to occupy for the electrons. We can think of the pole-skipping point as the transiting point where the fermi surface overlaps with a surface where electrons are tightly bound. In terms of physical properties, at that point, conducting and insulating phases overlap. Therefore, it could be interesting to investigate those systems from the perspective of holographic P-S phenomena. This we leave for our future studies.

\begin{acknowledgments}
We would like to acknowledge Wadbor Wahlang for his useful suggestions. BB would like to acknowledge the MHRD, Govt. of India for providing the necessary funding and fellowship to pursue research work.
\end{acknowledgments}

\onecolumngrid
\appendix
\begin{appendix}
\section{Dirac equations for various couplings}\label{appA}
    As discussed in Section 3, with various forms of coupling, we can write the 
 Dirac component equations. Imposing the spinor decomposition mentioned in the paper, we can get the component equations.
     \begin{itemize}
    \item \textbf{Dipole coupling :} When $\zeta(\phi)=-ip\phi\slashed{F}$ (here, $\slashed{F}=\frac{1}{2}\Gamma^{ab}e^{M}_{a}e^{N}_{b}F_{MN}$), then summing over the indices, we get, $\zeta(\phi)=-ip\phi(r) A_{v}'(r)\Gamma^{\underline{v}}$. 
\begin{subequations}
\begin{align}
&\notag r^{2}f(r)\partial_{r}\psi_{+}^{+}+\Gamma^{\underline{v}}\left[-i\omega+\frac{r^{2}f'(r)}{4}-iqA_{v}(r)+\frac{mr(1-f(r))}{2}+\frac{ik(1+f(r))}{2}-\frac{ipr\phi(r) A_{v}'(r)(1+f(r))}{2}\right]\psi_{-}^{-}\\&\label{a'} +\left[-i\omega+\frac{r^{2}f'(r)}{4}+\frac{3rf(r)}{2}-iqA_{v}(r)-\frac{mr(1+f(r))}{2}-\frac{ik(1-f(r))}{2}+\frac{ipr\phi(r) A_{v}'(r)(1-f(r))}{2}\right]\psi_{+}^{+}=0\\
 &\notag r^{2}f(r)\partial_{r}\psi_{-}^{-}-\Gamma^{\underline{v}}\left[-i\omega+\frac{r^{2}f'(r)}{4}-iqA_{v}(r)-\frac{mr(1-f(r))}{2}-\frac{ik(1+f(r))}{2}+\frac{ipr\phi(r) A_{v}'(r)(1+f(r))}{2}\right]\psi_{+}^{+}\\&\label{b'} +\left[-i\omega+\frac{r^{2}f'(r)}{4}+\frac{3rf(r)}{2}-iqA_{v}(r)+\frac{mr(1+f(r))}{2}+\frac{ik(1-f(r))}{2}-\frac{ipr\phi(r) A_{v}'(r)(1-f(r))}{2}\right]\psi_{-}^{-}=0
\end{align}
\end{subequations}
We can easily get the other two component equations for $\psi_{+}^{-}$ and $\psi_{-}^{+}$ by replacing $k\rightarrow -k$ in $\eqref{a'}$ and $\eqref{b'}$. 
\item \textbf{Yukawa coupling :} When $\zeta(\phi)=g\phi(r)$, then, the Dirac component equations we get as,
\end{itemize}
\begin{subequations}
\begin{align}
&\notag r^{2}f(r)\partial_{r}\psi_{+}^{+}+\Gamma^{\underline{v}}\left[-i\omega+\frac{r^{2}f'(r)}{4}-iqA_{v}(r)+\frac{mr(1-f(r))}{2}+\frac{ik(1+f(r))}{2}+\frac{gr\phi(r)(1-f(r))}{2}\right]\psi_{-}^{-}\\&\label{a1'} +\left[-i\omega+\frac{r^{2}f'(r)}{4}+\frac{3rf(r)}{2}-iqA_{v}(r)-\frac{mr(1+f(r))}{2}-\frac{ik(1-f(r))}{2} -\frac{gr\phi(r)(1+f(r))}{2}\right]\psi_{+}^{+}=0\\
&\notag r^{2}f(r)\partial_{r}\psi_{-}^{-}-\Gamma^{\underline{v}}\left[-i\omega+\frac{r^{2}f'(r)}{4}-iqA_{v}(r)-\frac{mr(1-f(r))}{2}-\frac{ik(1+f(r))}{2}-\frac{gr\phi(r)(1-f(r))}{2}\right]\psi_{+}^{+}\\&\label{b1'} +\left[-i\omega+\frac{r^{2}f'(r)}{4}+\frac{3rf(r)}{2}-iqA_{v}(r)+\frac{mr(1+f(r))}{2}+\frac{ik(1-f(r))}{2}+\frac{gr\phi(r)(1+f(r))}{2}\right]\psi_{-}^{-}=0
\end{align}
\end{subequations}
We can easily get the other two component equations for $\psi_{+}^{-}$ and $\psi_{-}^{+}$ by replacing $k\rightarrow -k$ in $\eqref{a1'}$ and $\eqref{b1'}$.
\section{Detailed pole-skipping analysis for Dipole coupling (real) case}\label{app-B}
     To calculate the pole-skipping points, we expand the spinors around the horizon as,
\begin{align}
\psi_{+}^{+}=\sum_{j=0}^{\infty}(\psi_{+}^{+})^{j}(r-r_0)^j,\hspace{1cm}\psi_{-}=\sum_{j=0}^{\infty}(\psi_{-}^{-})^{j}(r-r_0)^j
\end{align}
We have to expand the gauge field $A_v,f(r)$ and $\phi$ also around the horizon as\footnote{the emblackening factor $f(r)$ and gauge field $A_{v}(r)$ vanishes at the horizon.} 
\begin{align}
\begin{cases}
f(r)&=f(r_0)+(r-r_0)f'(r_0)+...\\
A_v(r)&=A_{v}(r_0)+(r-r_0)A'_v(r_0)+...\\
\phi(r)&=\phi(r_0)+(r-r_0)\phi^{'}(r_0)+...
\end{cases}
\end{align}

Expanding the Dirac equations $\eqref{a'}$ and $\eqref{b'}$  around the horizon as,
\begin{equation}
\mathcal{D}_{+}^{+}=\sum_{j=0}^{\infty}(\mathcal{D}_{+}^{+})^{j}(r-r_0)^j,\hspace{1cm}\mathcal{D}_{-}^{-}=\sum_{j=0}^{\infty}(\mathcal{D}_{-}^{-})^{j}(r-r_0)^j
\end{equation}

To calculate the zeroth order pole-skipping points, we expand $\eqref{a'}$ near horizon upto zeroth order and get,
\begin{subequations}
\begin{align}
\notag  (\mathcal{D}_{+}^{+})^{(0)}=&\left(-i\omega+\frac{{r_{0}}^{2}f'(r_0)}{4}+\frac{mr_0}{2}+\frac{ik}{2}-\frac{ipr_{0}\phi(r_0)A'_{v}(r_0)}{2}\right)(\psi_{-}^{-})^{(0)}+\left(-i\omega+\frac{{r_{0}}^{2}f'(r_0)}{4}\right.\\&\left.\hspace{4cm}-\frac{mr_0}{2}-\frac{ik}{2}+\frac{ipr_{0}\phi(r_0)A'_{v}(r_0)}{2}\right)(\psi_{+}^{+})^{(0)}=0
\end{align}
In the same way, we expand $\eqref{b'}$ upto zeroth order and get,
\begin{align}
\notag  (\mathcal{D}_{-}^{-})^{(0)}=&-\left(-i\omega+\frac{{r_{0}}^{2}f'(r_0)}{4}-\frac{mr_0}{2}-\frac{ik}{2}+\frac{ipr_{0}\phi(r_0)A'_{v}(r_0)}{2}\right)(\psi_{+})^{(0)}+\left(-i\omega+\frac{{r_{0}}^{2}f'(r_0)}{4}\right.\\&\left.\hspace{4cm}+\frac{mr_0}{2}+\frac{ik}{2}-\frac{ipr_{0}\phi(r_0)A'_{v}(r_0)}{2}\right)(\psi_{-})^{(0)}=0
\end{align}
\end{subequations}
To calculate the pole-skipping points, we make the coefficients of $(\psi_{+}^{+})^{(0)}$ and $(\psi_{-}^{-})^{(0)}$ zero and extract the frequency and associated momentum as,
\begin{equation}\label{ps}  
\omega_{0}=-\pi iT,\hspace{1cm}k=imr_{0}+p\mu\phi(r_0)
\end{equation}
So, this $(\omega_0,k)$ is the zeroth order pole-skipping point. We will get another pole-skipping point from the other two component equations simply by replacing $k\rightarrow -k$.
\begin{equation} 
\omega_{0}=-\pi iT,\hspace{1cm}k=-imr_{0}-p\mu\phi(r_0)
\end{equation}
The additional term we got is dependent on coupling parameter $p$, a chemical potential $\mu$ and the value of the scalar field at the horizon $\phi(r_0)$. We have discussed the effect of this additional term in section 3 of the paper.\par
 To get the higher-order pole-skipping points, we have to expand the 1st-order Dirac equations up to higher orders. For that, we will follow a matrix formalism to extract the pole-skipping points. Now, expanding the Dirac equation in the first order, we will get two equations containing the coefficients of $(\psi_{+}^{+})^{1},(\psi_{-}^{-})^{1},(\psi_{+}^{+})^{0}$ and $(\psi_{-}^{-})^{0}$. Then, we will form a $2\times 2$ matrix of the coefficients. We can write the equations in a matrix form as,
\begin{equation}
\begin{pmatrix}
(\mathcal{D}_{+}^+)^{1}\\(\mathcal{D}_{-}^-)^{1}
\end{pmatrix}=\mathcal{\tilde{A}}^{11}\begin{pmatrix}
(\psi_{+}^{+})^{1}\\(\psi_{-}^{-})^{1}
\end{pmatrix}+\mathcal{\tilde{A}}^{10}\begin{pmatrix}
(\psi_{+}^{+})^{0}\\(\psi_{-}^{-})^{0}
\end{pmatrix}=0
\end{equation}
Where,\begin{equation}
\mathcal{\tilde{A}}^{11}=\begin{pmatrix}
5\pi T-i\omega-\frac{mr_0}{2}-\frac{ik}{2}+\frac{ipr_{0}\phi(r_0)A'_{v}(r_0)}{2} & \left(-i\omega+\pi T+\frac{mr_0}{2}+\frac{ik}{2}-\frac{ipr_{0}\phi(r_0)A'_{v}(r_0)}{2} \right)\\-\left(-i\omega+\pi T-\frac{mr_0}{2}-\frac{ik}{2}+\frac{ipr_{0}\phi(r_0)A'_{v}(r_0)}{2} \right)& 5\pi T-i\omega+\frac{mr_0}{2}+\frac{ik}{2}-\frac{ipr_{0}\phi(r_0)A'_{v}(r_0)}{2}
\end{pmatrix}
\end{equation}
and, $\mathcal{\tilde{A}}^{10}$ is dependent on $k$ only. We can clearly see that det$ \,\mathcal{\tilde{A}}^{11}=8\pi T(3\pi T-i\omega)$ which disappears at $\omega=-3\pi iT $. To calculate associated momentum, we have to rewrite  $(\mathcal{D}_{+}^+)^{0},(\mathcal{D}_{+}^+)^{1}$ and $(\mathcal{D}_{-}^{-})^{1}$ at $\omega=-3\pi iT $ and we can construct a $3\times 3 $ matrix out of the co-efficients of $(\psi_{+}^{+})^{0},(\psi_{-}^{-})^{1}$ and $\psi_{c}^{1}$ where $\psi_{c}^{1}=(\psi_{+}^{+})^{1}-\Gamma^{\underline{v}}(\psi_{-}^{-})^{1}$. We can express the matrix form as,
\begin{equation}\label{q} 
\begin{pmatrix}
(\mathcal{D}_{+}^+)^{0}\\(\mathcal{D}_{+}^+)^{1}\\(\mathcal{D}_{-}^-)^{1}
\end{pmatrix}=\mathcal{\tilde{A}}_1(\omega_1,k)\begin{pmatrix}
(\psi_{+}^{+})^{0}\\(\psi_{-}^{-})^{0}\\\psi_{c}^1\end{pmatrix}=\begin{pmatrix}
\mathcal{\tilde{A}}_{++}^{00} & \mathcal{\tilde{A}}_{+-}^{00} & 0\\\mathcal{\tilde{A}}_{++}^{10} & \mathcal{\tilde{A}}_{+-}^{10} & \mathcal{\tilde{A}}_{+}^{11}\\\mathcal{\tilde{A}}_{-+}^{10} & \mathcal{\tilde{A}}_{--}^{10} & \mathcal{\tilde{A}}_{-}^{11}
\end{pmatrix}\begin{pmatrix}
\psi_{+}^0\\\psi_{-}^0\\\psi_{c}^1\end{pmatrix}=0
\end{equation}
We can evaluate the associated momentum at which det$\mathcal{\tilde{A}}_1(\omega_1,k)$ vanishes. To see the associated momentum with $\omega=-3\pi iT$, we evaluate the above matrix $\eqref{q}$ at $\omega=-3\pi iT$ 
In this way, we have extracted the pole-skipping points up to many orders. We can clearly see that these $\omega$'s are fermionic matsubara frequencies at various orders; $\omega=\omega_{n}=-2\pi iT(n+\frac{1}{2})$. 
\begin{align}
\omega_{1}&=-3\pi i T\\
k_{1a}&=\frac{ir_0}{3 T}\left(2^{-1/3}A+T\left(m-\frac{3ip\mu\phi(r_0)}{r_0}\right)+\frac{4^{2/3}T^{2}B}{r_{0}A}\right)\\
k_{1b}&=-\frac{ir_0}{3T}\left(4^{-2/3}(1-\sqrt{3i})A-T\left(m-\frac{3ip\mu\phi(r_0)}{r_0}\right)+\frac{2^{1/3}(1+\sqrt{3i})T^2B}{r_0 A}\right)\\
k_{1c}&=-\frac{ir_0}{3T}\left(4^{-2/3}(1+\sqrt{3i})A-T\left(m-\frac{3ip\mu\phi(r_0)}{r_0}\right)+\frac{2^{1/3}(1-\sqrt{3i})T^2 B}{r_0 A}\right)\\
B&=(-9+2m^2)r_{0}+18\pi T+3iq\mu\\\notag
A&=r_{0}^{-2/3}\left[-12 mr_{0}B T^{3}-2m^{2}r_{0}-9\pi T-\frac{27}{4}ip\mu\phi(r_0)r_{0}T^{3}(m_{\phi}^{2}+\phi(r_0)^2-4\pi T)\right.\\&\left.+4ir_{0}^{1/2}T^{3}\left(2B^{3}-B m+2m^{3}r_{0}+9m\pi T-\frac{27}{4}ip\mu\phi(r_0)T^{3}(m_{\phi}^{2}+\phi(r_0)^2-4\pi T)\right)^{1/2}\right]^{1/3}
\end{align}
So, these are the 1st order pole-skipping points. we get 3 associated momentum values in 1st order. We can see from the momentum values that they are directly proportional to $\phi(r_0)$ which means $(T_c-T)^{1/2}$. So, up to higher orders, they obey the power law. From the \figurename{\ref{yuk_dip_real_plot}} also, it is evident.
 \section{Higher order momentum values for Yukawa coupling (real)}\label{app-c}
 With the same method described above, we can calculate the 1st-order P-S points for Yukawa coupling. We observe that, $\omega_1=-3\pi iT$ while $k_1$ takes 3 values as
 \begin{subequations}
\begin{align}
\omega_{1}&=-3\pi iT\\
k_{1a}&=\frac{ir_0}{3T}\left(\frac{Z}{r_0}+(m+g\phi(r_0))T+\frac{2Y}{Z}T^2\right)\\
k_{1b}&=\frac{ir_0}{3T}\left(-\frac{(1-\sqrt{3}i)Z}{2r_0}+(m+g\phi(r_0))T-\frac{(1+\sqrt{3}i)Y}{Z}T^2\right)\\
k_{1c}&=\frac{ir_0}{3T}\left(-\frac{(1+\sqrt{3}i)Z}{2r_0}+(m+g\phi(r_0))T-\frac{(1-\sqrt{3}i)Y}{Z}T^2\right)\\
Y&=(-9+2m^2)r_{0}+18\pi T+3iq\mu+4gmr_{0}\phi(r_0)+2g^{2}r_{0}\phi^2(r_0)\\
\notag Z&=\left(2mr_{0}^{2}T^{3}(Y+3(12r_{0}-2m^{2}r_0-15\pi T-4iq\mu)-16gmr_{0}\phi(r_0)-14g^{2}r_{0}\phi^2(r_0)\right.\\\notag&\left.+27\frac{g}{m}\phi(r_0)(r_{0}-\pi T)-9i\frac{g}{m}q\mu\phi(r_0)-4\frac{g^3}{m}r_{0}\phi^3(r_0)+\frac{27g}{4m}r_{0}\phi(r_0)(m_{\phi}^2+\phi^2(r_0)))\right.\\\notag&\left.+\frac{1}{2}r_{0}^{3/2}T^{3}(-32 Y^{3}+r_{0}\left(4m(r_0(-27+4m^2)+27\pi T+9iq\mu)+3g(4(-9r_{0}+4m^{2}r_{0}+9\pi T+3iq\mu)\right.\right.\\&\left.\left.-9r_{0}m_{\phi}^2)\phi(r_0)+48g^{2}mr_{0}\phi^2(r_0)+g(-27+16g^2)r_{0}\phi^3(r_0)\right)^2) \right)^{1/3}
\end{align}
\end{subequations}
The momentum values are proportional to $\phi(r_0)$ here also. So, they will obey the power law as $(T_c-T)^{1/2}$. So, we have checked that up to the first order, P-S points obey $(T_c-T)^{1/2}$ law near critical temperature for a theory with massless fermions propagating in a charged black hole background, with real dipole and Yukawa coupling. We have explicitly discussed this in subsection 3.1.1.
\section{Dirac component equations for charged Dipole coupling}\label{app-d}
With charged Dipole coupling, decomposing the Dirac equation into components, we can write,
\begin{subequations}
\begin{align}
&\notag r^{2}f(r)\partial_{r}\psi_{+}^{+}+\Gamma^{\underline{v}}\left[-i\omega+\frac{r^{2}f'(r)}{4}-iqA_{v}(r)+\frac{mr(1-f(r))}{2}+\frac{ik(1+f(r))}{2}-\frac{ipr|\tilde{\phi}|^{2} A_{v}'(r)(1+f(r))}{2}\right]\psi_{-}^{-}\\&\label{aa'} +\left[-i\omega+\frac{r^{2}f'(r)}{4}+\frac{3rf(r)}{2}-iqA_{v}(r)-\frac{mr(1+f(r))}{2}-\frac{ik(1-f(r))}{2}+\frac{ipr|\tilde{\phi}|^{2} A_{v}'(r)(1-f(r))}{2}\right]\psi_{+}^{+}=0\\
 &\notag r^{2}f(r)\partial_{r}\psi_{-}^{-}-\Gamma^{\underline{v}}\left[-i\omega+\frac{r^{2}f'(r)}{4}-iqA_{v}(r)-\frac{mr(1-f(r))}{2}-\frac{ik(1+f(r))}{2}+\frac{ipr|\tilde{\phi}|^{2} A_{v}'(r)(1+f(r))}{2}\right]\psi_{+}^{+}\\&\label{bb'} +\left[-i\omega+\frac{r^{2}f'(r)}{4}+\frac{3rf(r)}{2}-iqA_{v}(r)+\frac{mr(1+f(r))}{2}+\frac{ik(1-f(r))}{2}-\frac{ipr|\tilde{\phi}|^{2} A_{v}'(r)(1-f(r))}{2}\right]\psi_{-}^{-}=0
 \end{align}
 \end{subequations}
 The other two component equations can be obtained by replacing $k\rightarrow -k$ in the above two equations. Expanding these equations near the black hole horizon and applying the method discussed in \eqref{app-B}, we can calculate P-S points up to any order. We have checked that the behaviour of momentum values for this case is proportional to $(T_c-T)$ which is different from the real coupling case. We have calculated the P-S points in section 3.2. The $\omega$ values are not different from the fermionic Matsubara frequencies, but the momentum values change with the coupling. We have shown the numerical behaviour of these higher-order points in Figure 5.
 \section{Dirac component equations for charged Yukawa coupling}\label{app-e}
 With charged Yukawa coupling, decomposing the Dirac equation into components, we can write,
\begin{subequations}
\begin{align}
&\notag r^{2}f(r)\partial_{r}\psi_{+}^{+}+\Gamma^{\underline{v}}\left[-i\omega+\frac{r^{2}f'(r)}{4}-iqA_{v}(r)+\frac{mr(1-f(r))}{2}+\frac{ik(1+f(r))}{2}+\frac{gr|\tilde{\phi}|^{2}(1-f(r))}{2}\right]\psi_{-}^{-}\\&\label{e1'} +\left[-i\omega+\frac{r^{2}f'(r)}{4}+\frac{3rf(r)}{2}-iqA_{v}(r)-\frac{mr(1+f(r))}{2}-\frac{ik(1-f(r))}{2} -\frac{gr|\tilde{\phi}|^{2}(1+f(r))}{2}\right]\psi_{+}^{+}=0\\
&\notag r^{2}f(r)\partial_{r}\psi_{-}^{-}-\Gamma^{\underline{v}}\left[-i\omega+\frac{r^{2}f'(r)}{4}-iqA_{v}(r)-\frac{mr(1-f(r))}{2}-\frac{ik(1+f(r))}{2}-\frac{gr|\tilde{\phi}|^{2}(1-f(r))}{2}\right]\psi_{+}^{+}\\&\label{e2'} +\left[-i\omega+\frac{r^{2}f'(r)}{4}+\frac{3rf(r)}{2}-iqA_{v}(r)+\frac{mr(1+f(r))}{2}+\frac{ik(1-f(r))}{2}+\frac{gr|\tilde{\phi}|^{2}(1+f(r))}{2}\right]\psi_{-}^{-}=0
\end{align}
\end{subequations}
 The other two component equations can be obtained by replacing $k\rightarrow -k$ in the above two equations. With the matrix method, we can calculate the P-S points. Here also, $\omega$ values are just the fermionic Matsubara frequencies, only the momentum value changes. The behaviour of momentum values for this case is also proportional to $(T_c-T)$.
\end{appendix}
\twocolumngrid
\bibliographystyle{apsrev4-1}
\bibliography{epjc}
\end{document}